\documentclass[journal=jacsat,manuscript=article]{achemso}

\usepackage[T1]{fontenc} 
\usepackage{graphicx}  
\usepackage{dcolumn}   
\usepackage{bm}        
\usepackage{verbatim}   
\usepackage{graphicx}
\usepackage{epstopdf}
\usepackage{footnote}
\usepackage{epsfig}
\usepackage{lipsum}
\usepackage{subfigure}
\usepackage{amssymb}
\usepackage{amsmath,bm}
\usepackage{xcolor}
\usepackage{float,natbib}
\usepackage{subfigure}
\usepackage[most]{tcolorbox}
\usepackage{soul}
\usepackage{color}

\author{Sajjad Taravati}
\affiliation{Current address: The Edward S. Rogers Sr. Department of Electrical and Computer Engineering, University of Toronto, Toronto, Ontario M5S 3H7, Canada}
\email{sajjad.taravati@utoronto.ca}
\author{George V. Eleftheriades}

\title[An \textsf{achemso} demo]
  {Highly Linear Nonmagnetic Circulator Enabled By A\\ Temporal Nonreciprocal Phase Shifter}

\abbreviations{IR,NMR,UV}

\begin{document}


\begin{abstract}
Conventional circulators are made of magnetic ferrites and suffer from a cumbersome architecture, incompatibility with integrated circuit technology and inability for high frequency applications. To overcome these limitations, here we propose a lightweight low-profile non-magnetic circulator comprising a nonreciprocal time-varying phase shifter. This circulator is composing a nonreciprocal temporal phase shifter and two reciprocal delay-line-based phase shifters. The proposed nonreciprocal temporal phase shifter is based on the generation of time-harmonic signals, enforcing destructive interference for undesired time harmonics and constructive interference for desired time harmonics at different locations of the structure. Such a unique task is accomplished through two phase-engineered temporal loops. The phase and frequency of these two loops are governed by external signals with different phases, imparting an effective electronic angular momentum to the system. We observe large isolation level of greater than 32 dB, a P1dB of +31.7 dBm and IIP3 of +42.4 dBm. Furthermore, this circulator is endowed with a reconfigurable architecture and can be directly embedded in a conventional integrated circuit (IC) technology to realize a class of high power handling and linear IC circulators.
\end{abstract}

\section{Introduction}
Circulators are three port nonreciprocal devices and represent key elements for full-duplex telecommunication transceivers~\cite{ohm1956broad,fay1965operation,linkhart2014microwave}. They are used to separate the transmitter and receiver paths by introducing nonreciprocity between their ports~\cite{ewing1967ring,linkhart2014microwave}. Conventional circulators are formed by ferrite magnetic materials, where nonreciprocity is achieved under a magnetic bias field~\cite{fay1965operation,o2009experimental,fan2012magnetically}. However, ferrite-based magnetic circulators are bulky, costly, heavy and are not available at high frequencies. Recently, nonmagnetic nonreciprocity has spurred a surge of research activity to eliminate the issues associated with magnetic nonreciprocal devices.

Nonmagnetic nonreciprocity is conventionally achieved through nonreciprocal properties of transistors at microwave frequencies~\cite{tanaka1965active,smith1988gaas,ayasli1989field,kalialakis2000analysis,carchon2000power,taravati2021programmable,taravati2021full}, and electro-optical modulators at optical frequencies~\cite{bhandare2005novel}. However, these techniques substitute the absence of cumbersome magnetic bias for other tangible drawbacks, such as poor noise performance and strong nonlinearities of transistors,
or the complexity and large size of the electro-optical modulators. Recently, space-time modulation has been shown to be a remarkable approach to realize nonreciprocal components, especially for integrated optical networks where it can be fully realized using silicon photonics~\cite{Fan_NPH_2009,Taravati_Kishk_MicMag_2019,camacho2020achieving}. It is shown that by taking advantage of one-way progressive wave coupling in a space-time-modulated structure, nonreciprocal devices and isolators can be created~\cite{Fan_NPH_2009,Taravati_PRB_2017,Taravati_PRB_SB_2017,li2019nonreciprocal,Taravati_PRAp_2018,Harry_Atwater_time,taravati_PRApp_2019}. The space-time modulation technique exhibits appealing capabilities, such as parametric wave amplification~\cite{tien1958traveling,tien1958parametric,zhu2020tunable}, unidirectional beam splitting~\cite{Taravati_Kishk_PRB_2018}, pure frequency generation~\cite{Taravati_PRB_Mixer_2018,taravati2021pure}, nonreciprocal metasurfaces~\cite{zang2019nonreciprocal_metas,Taravati_Kishk_TAP_2019,taravati2020four}, antennas~\cite{zang2019nonreciprocal}, nonreciprocal filters~\cite{wu2020frequency}, and full-transceiver realization~\cite{Taravati_LWA_2017,Taravati_AMA_PRApp_2020}.

Nevertheless, the space-time modulation technique requires optically long structures as they rely on the weak and progressive wave coupling, acousto-optic and electro-optic effects~\cite{huang2011complete,Fan_PRL_109_2012,Taravati_PRB_2017,Taravati_PRB_SB_2017,Taravati_Kishk_TAP_2019}. Time-varying circulators~\cite{estep2014magnetic,dinc2017millimeter,kord2017magnet,nagulu2018nonreciprocal,Taravati_Kishk_MicMag_2019,kadry2020angular} have shown to be a good candidate to realize versatile nonmagnetic wave circulators. They offer outstanding noise performance, and are compatible with integrated circuit technology and high frequency applications. Additionally, phase-engineered time modulation has recently been considered as an alternative approach to realize unique functionalities and devices~\cite{zang2019nonreciprocal,taravati2020full}. Phase-engineered time modulated structure provide a low-profile and efficient platform for wave-engineering by taking advantage of asymmetric frequency-phase transitions.

Within this context, this study proposes a low-profile, lightweight and nonmagnetic circulator based on a nonreciprocal phase shifter at microwaves. This nonreciprocal phase shifter is constituted of two time-modulated loops, each of which comprising phase-engineered temporal transmission lines. The lightweight and low-profile properties originate from the fact that the time modulation requires only four temporal varactors. Isolation between different ports is achieved by imposing specified constructive and destructive interferences at the ports of the circulator. We show that by proper design of the band structure of the phase-engineered temporal transmission lines, a desired nonreciprocal phase shift and consequently, strong signal isolation between the three ports of the circulator can be achieved. The frequency and phase transition in a time-varying transmission line are provided by varactors biased by a time-varying voltage. The circulator exhibits strong isolation of more than 33 dB, highly linear response, that is, 1 dB compression point of 31.7 dBm, broad-band operation  where more than 20 dB isolation is introduced across a 19.2$\%$ fractional bandwidth. This circulator can be realized using integrated circuit (IC) technology for a much smaller dimension and programmable mechanism.

\section{Results}

\subsection{Circulator Design}

\begin{figure}
	\centering
	\includegraphics[width=0.7\columnwidth]{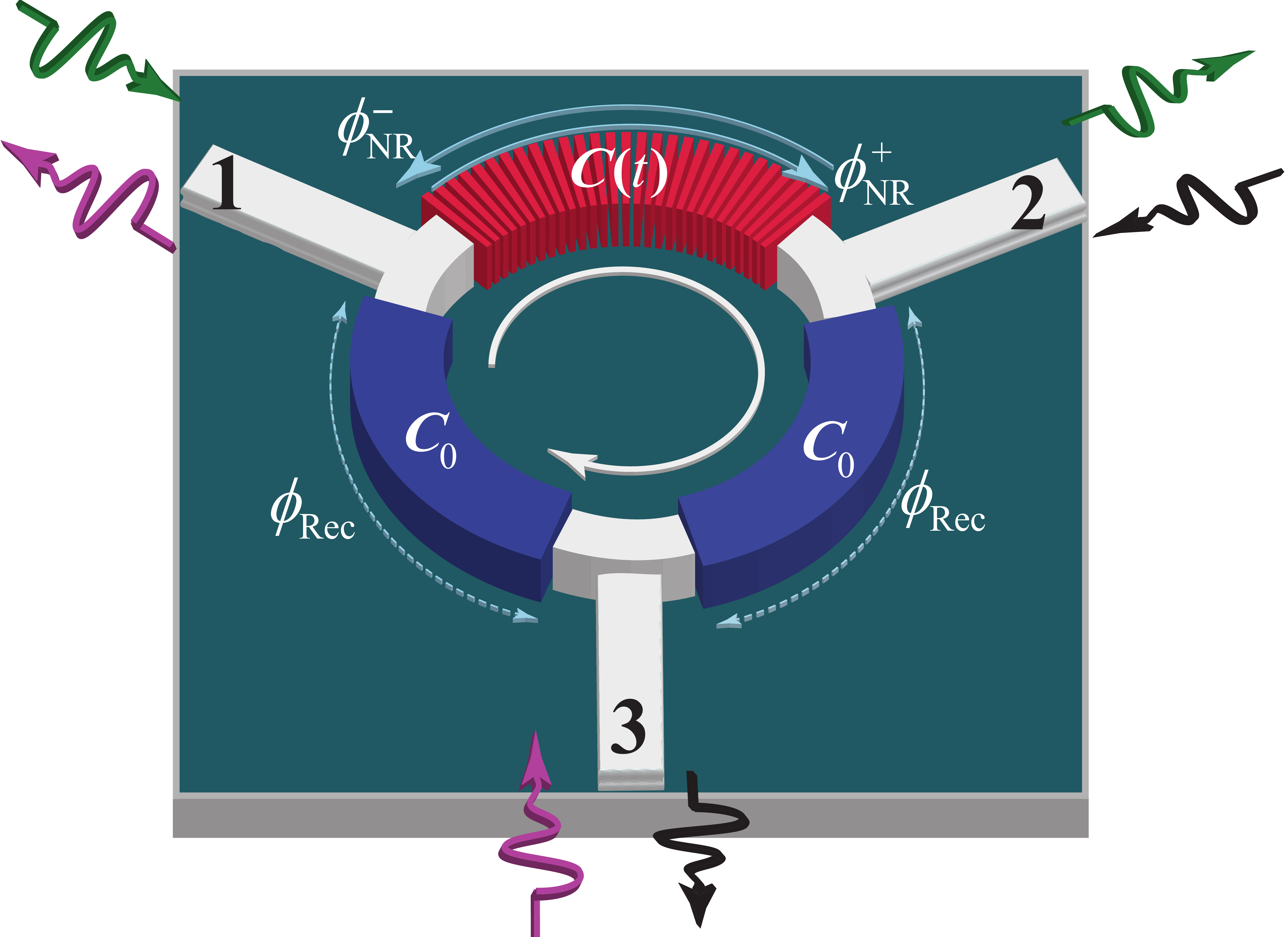}
	\caption{Schematic illustration of the time-modulated circulator composed of a nonreciprocal time-varying phase-shifter and two reciprocal phase shifters.}
	\label{fig:1}
\end{figure}

Figure~\ref{fig:1} depicts the architecture of the proposed time-modulated circulator. The circulator is composed of a nonreciprocal time-varying phase shifter between ports 1 and 2, and two reciprocal phase shifters (e.g., transmission lines) between ports 1 and 3 and between ports 2 and 3. To compute the required phase shifts at the three arms of the circulator, we excite a given port and ensure constructive interference at the next right-hand port, and destructive interference at the next left-hand port. For instance, we first excite port 1. Then, a full transmission (constructive interference) from port 1 to port 2 requires
\begin{subequations}
	\begin{equation}
		\phi_\text{NR}^+=2 \phi_\text{Rec},
	\end{equation}
	whereas a null (destructive interference) at port 3 may be achieved by satisfying
	\begin{equation}
		\phi_\text{Rec}=\phi_\text{NR}^+ +\phi_\text{Rec}-\pi,
	\end{equation}
	where $\phi_\text{NR}^+$ is the forward phase shift of the nonreciprocal phase shifter, from left to right, and $\phi_\text{Rec}$ is the phase shift provided by reciprocal phase shifters. Next, we excite port 2, where a full transmission (constructive interference) from port 2 to port 3 requires
	\begin{equation}
		\phi_\text{Rec}=\phi_\text{NR}^- +\phi_\text{Rec},
	\end{equation}
	where $\phi_\text{NR}^-$ denotes the backward phase shift of the nonreciprocal phase shifter, from right to left. A null (destructive interference) at port 1 can be satisfied by
	\begin{equation}
		\phi_\text{NR}^-=2 \phi_\text{Rec}-\pi.
	\end{equation}
\end{subequations}

Hence, by setting $\phi_\text{Rec}=\pi/2$, $\phi_\text{NR}^+=\pi$ and $\phi_\text{NR}^-=0$, the desired constructive and destructive interferences between three ports of the circulator in Fig.~\ref{fig:1} can be achieved.

The two reciprocal phase shifters may be realized by lumped elements at low frequencies or using transmission lines at high frequencies. The nonreciprocal phase shifter is realized by a time-varying lightweight and highly efficient phase shifter, where the effective electric permittivity of a transmission line is time modulated. 

\subsection{Temporal Nonreciprocal Phase shifter}

\begin{figure}
	\centering
	\includegraphics[width=1\columnwidth]{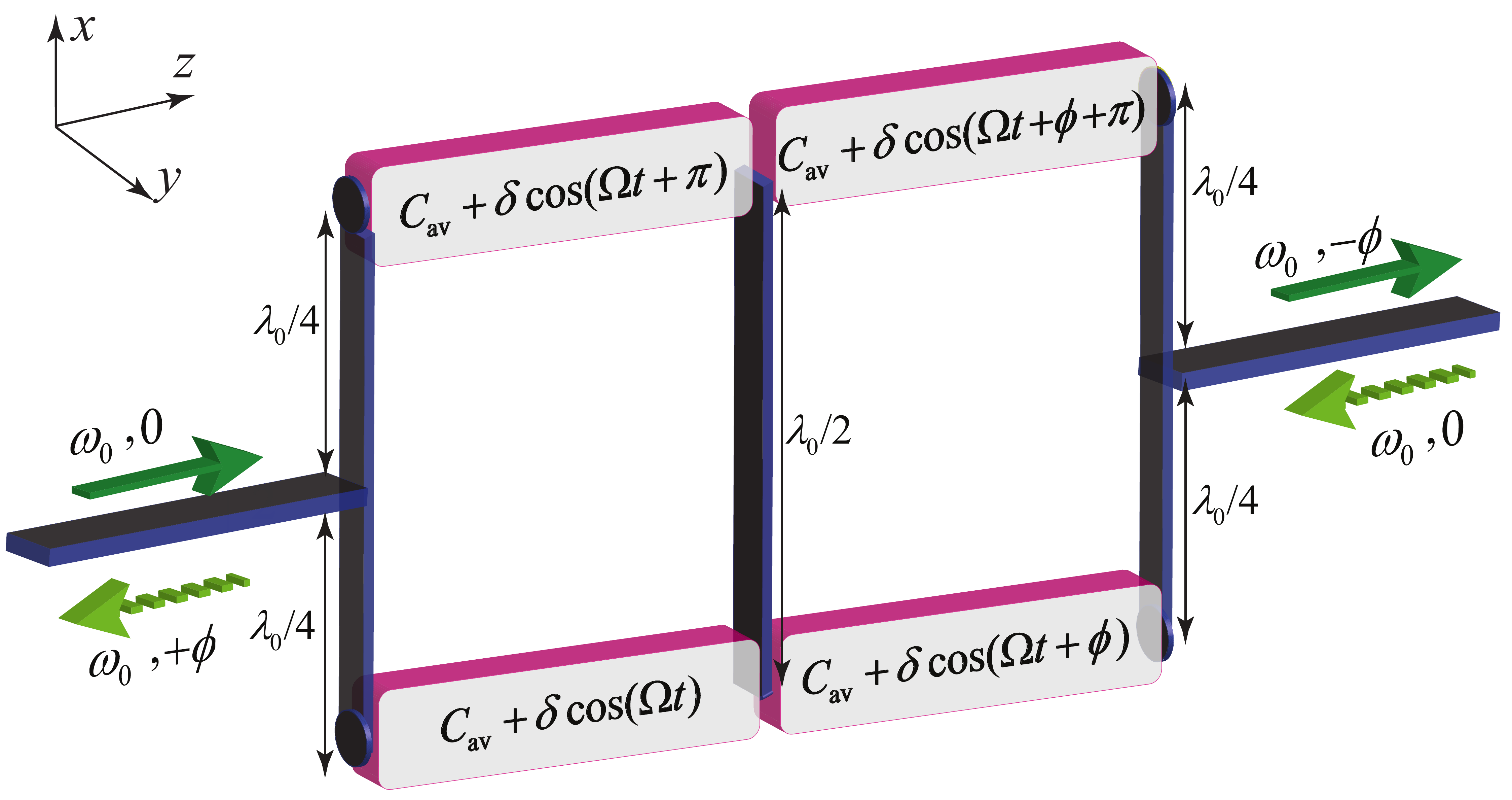}
	\caption{Schematic representation of the nonreciprocal phase shifter integrating two time-modulated loops.}
	\label{fig:sch}
\end{figure}

Figure~\ref{fig:sch} depicts the architecture of the temporal nonreciprocal phase shifter, formed by two time-varying loops, each of which is composed of two phase-engineered temporal transmission lines. The left temporal loop is composed of two (upper and lower) temporal transmission lines, where the capacitance of the transmission lines is modulated in time considering a $\pi$ phase difference between the upper and lower temporal transmission lines, i.e.,

\begin{subequations}\label{fig:capac}
	\begin{equation}
		C_1(t)=C_\text{av}+\delta \cos(\Omega t),
	\end{equation}
	\begin{equation}
		C_1'(t)=C_\text{av}+\delta \cos(\Omega t+\pi),
	\end{equation}
	where $C_\text{av}$ is the average capacitance of the loops and $\delta$ is the modulation amplitude of the loops. In addition, $\Omega$ denotes the temporal modulation frequency. The right temporal loop is composed of two (upper and lower) temporal transmission lines, where the capacitance of the transmission lines is modulated in time considering a $\pi$ phase difference between the upper and lower temporal transmission lines and $\phi$ phase difference with the left loop, i.e., 	
	\begin{equation}
		C_2(t)=C_\text{av}+\delta \cos(\Omega t+\phi),
	\end{equation}
	\begin{equation}
		C_2'(t)=C_\text{av}+\delta \cos(\Omega t+\phi+\pi),
	\end{equation}
\end{subequations}
where $\phi$ represents the modulation phase difference between the first and second loops. Next, by proper design of the time modulation of the two loops, a desired nonreciprocal phase shift is achieved. These two temporal loops are designed to provide appropriate constructive interference for the desired time harmonics and destructive interference for the undesired time harmonics. 

\begin{figure}
	\centering
	\includegraphics[width=0.7\columnwidth]{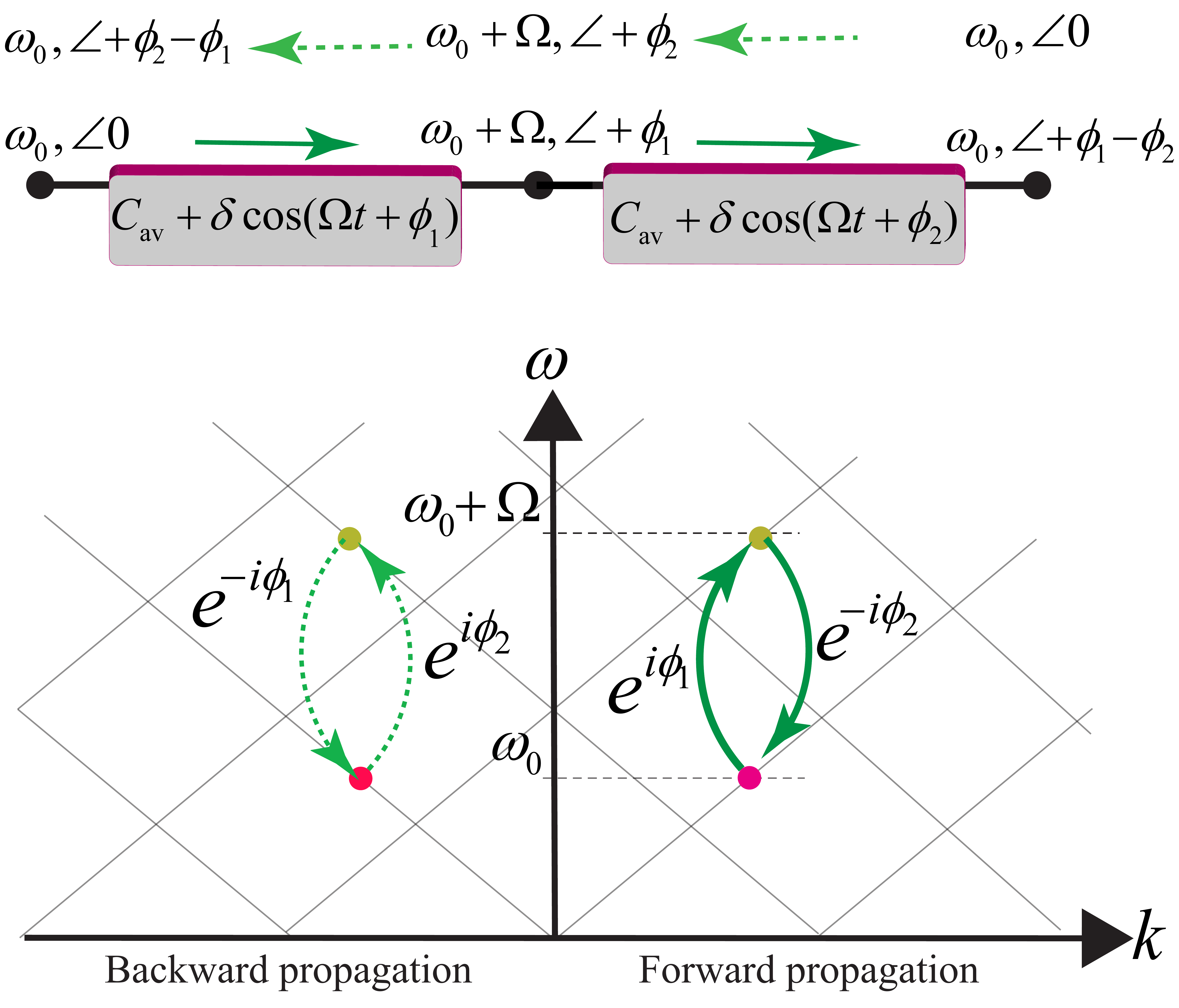}
	\caption{Schematic representation of forward and backward frequency-phase transitions in two cascaded temporal transmission line.}
	\label{fig:disp2}
\end{figure}

Figure~\ref{fig:disp2} shows a schematic representation of the dispersion diagram for asymmetric frequency-phase transitions in a temporal transmission line. Here, a frequency up-conversion from $\omega_0$ to $\omega_0+\Omega$ is accompanied by a positive additive phase, i.e., $+\phi_1$ in the forward transmission and $+\phi_2$ in the backward transmission. However, a frequency down-conversion from $\omega_0+\Omega$ to $\omega_0$ is accompanied by a negative additive phase, i.e., $-\phi_2$ in the forward transmission and $-\phi_1$ in the backward transmission.

To best understand the effect of the phase and amplitude of the time modulation on the output voltage of the temporal transmission line, we first investigate variation of the voltage through a temporal-varactor-loaded transmission line. It may be shown that, by proper engineering of the dispersion of the transmission line, the output voltage acquires a frequency transition (up/down-conversion) accompanied by a phase transition. We assume the dispersion bands of the structure are engineered so that only the fundamental and the first higher order time harmonics are excited, whereas all higher order time harmonics are suppressed. Hence, the voltage is defined based on the superposition of the $n=0$ and $n=-1$ space-time harmonics fields, i.e.,
\begin{equation}\label{eqa:el}
	V_\text{S}(z,t)=v_{0}(z) e^{-i \left(k_z z -\omega_0 t \right)}+v_{1}(z) e^{-i \left(k_z z -(\omega_0+\Omega) t \right)}.
\end{equation}

Equation~\eqref{eqa:el} considers an identical wave number $k_0$ for both the fundamental and the higher-order harmonics as the transmission line is modulated only in time, and not in space. It may be seen from Fig.~\ref{fig:disp2} that, as a result of the temporal sinusoidally periodic capacitance, the dispersion diagram of the temporal transmission line is periodic with respect to the $\omega$ axis~\cite{Halevi_PRA_2009}, that is $k(\omega_0+\Omega)=k(\omega_0)=k_0$. Such a temporal periodicity leads to vertical electromagnetic transitions in the dispersion diagram, as shown in Fig.~\ref{fig:disp2}. The unknown spatially variant amplitudes $v_{0}(z)$ and $v_{1}(z)$ are to be found through satisfying both the Telegrapher’s equations and initial conditions at $z=0$. We then consider a lossless transmission line with time-varying capacitance, the Telegrapher’s
equations read%
\begin{subequations}
	\begin{equation}\label{eqa:T1}
		\frac{\partial V(z,t) }{\partial z}=-L_0 \frac{\partial I(z,t) }{\partial t},
	\end{equation}
	\begin{equation}\label{eqa:T2}
		\frac{\partial I(z,t) }{\partial z}=-C_0 \frac{\partial [ C_\text{eq}(t) V(z,t)] }{\partial t},
	\end{equation}
\end{subequations}
where $L_0$ and $C_0$ are the inductance and capacitance per unit length of the transmission line,
respectively. Equations~\eqref{eqa:T1} and~\eqref{eqa:T2} yield
\begin{align}\label{eqa:wave_eq}
	\frac{\partial^{2} V(z,t)}{\partial z^{2}}= \dfrac{1}{v_\text{p}^2} \frac{\partial^{2} [C_\text{eq}(t) V(z,t)]}{\partial t^{2}},
\end{align}
where $v_\text{p}=1/\sqrt{L_0 C_0}$.

We then insert the voltage in~\eqref{eqa:el} into~\eqref{eqa:wave_eq} and take the space and time derivatives, while considering a slowly varying envelope approximation, and apply $\int_{0}^{2\pi/\Omega}dt$ to both sides of the resultant equation, which yields
\begin{equation}\label{eqa:eq222}
	\begin{split}
		\frac{d v_{0}(z)}{d z}   =  \frac{ik_0 \delta}{4C_\text{av}}  e^{-i\phi} v_{1}(z)   ,
	\end{split}
\end{equation}
\begin{equation}\label{eqa:eq222b}
	\begin{split}
		\frac{d v_{1}(z)}{d z}   =  \frac{ik_1^2 \delta}{4 k_z C_\text{av}}  e^{i\phi} v_{0}(z)   ,
	\end{split}
\end{equation}

Next, we seek independent differential equations for $a_0(z)$ and $v_{1}(z)$, which yields

\begin{align}\label{eqa:eq3}
	\frac{d^{2}v_{0}(z)}{dz^{2}}-\gamma_0 \gamma_1v_{0}(z)=0,\\
	\frac{d^{2}v_{1}(z)}{dz^{2}}-\gamma_0 \gamma_1v_{1}(z)=0.
\end{align}
which are second order differential equations, where $\gamma_0=ik_0 \delta e^{-i\phi}/4 C_\text{av}  $ and $\gamma_1=ik_1^2 \delta e^{i\phi}/4 k_z C_\text{av}$.

The up-conversion assumes the initial conditions of $v_{0}(0)=A_0$ and $v_{1}(0)=0$, which gives
\begin{subequations}
	\begin{equation}\label{eq:up1}
		v_{0}(z)=A_0 \cos\left(\dfrac{\delta k_1}{4 C_\text{av}} z\right),
	\end{equation}
	\begin{equation}\label{eq:up2}
		v_{1}(z)=A_0\dfrac{k_1 }{k_0} e^{+i \phi}  \sin\left(\dfrac{\delta k_1}{4 C_\text{av}} z\right).
	\end{equation}
\end{subequations}

Equations~\eqref{eq:up1} and~\eqref{eq:up2} show that the change of the frequency and phase of the up-converted signal corresponds to the frequency and phase of the temporal modulation signal, $\Omega$ and $+\phi$. In addition, the maximum amplitude of the up-converted signal $a_1(z)$ occurs at $\delta k_1/ C_\text{av}=2\pi$. This reveals that, one can control the amplitude of the up-converted signal by changing the modulation amplitude $\delta$. Following the same procedure, we can determine the input and output voltages for the down-conversion. 

The down-conversion assumes the initial conditions of $v_{0}(0)=0$ and $v_{1}(0)=A_1=\text{Max}[v_{1}(z)]=A_0 k_1 /k_0 \exp(+i \phi_0)$, which gives
\begin{subequations}
	\begin{equation}\label{eq:down1}
		v_{1}(z)=A_1 \cos\left(\dfrac{\delta k_1}{4 C_\text{av}} z\right),
	\end{equation}
	and
	\begin{equation}\label{eq:down2}
		v_{0}(z)= A_1\dfrac{k_0 }{k_1} e^{-i \phi}  \sin\left(\dfrac{\delta k_1}{4 C_\text{av}} z \right)=A_0  e^{i (\phi_0-\phi)}  \sin\left(\dfrac{\delta k_1}{4 C_\text{av}} z \right).
	\end{equation}
\end{subequations}

Equations~\eqref{eq:down1} and~\eqref{eq:down2} show that the change of the frequency and phase of the down-converted signal corresponds to the frequency and phase of the temporal modulation signal, $\Omega$ and $-\phi$.

\begin{figure}
	\begin{center}
		\subfigure[]{\label{fig:p1} 
			\includegraphics[width=0.7\columnwidth]{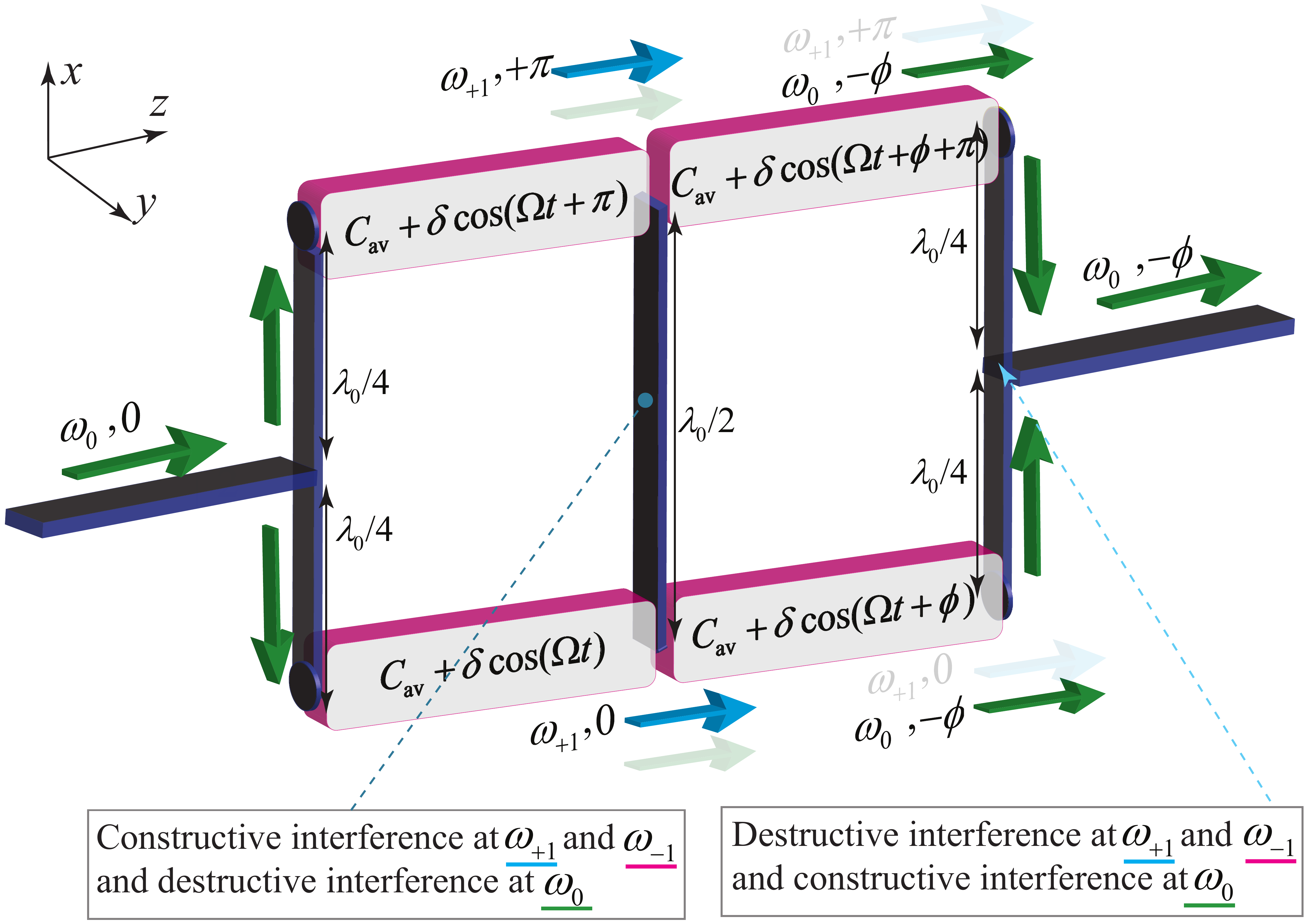}} 
		\subfigure[]{\label{fig:p2} 
			\includegraphics[width=0.7\columnwidth]{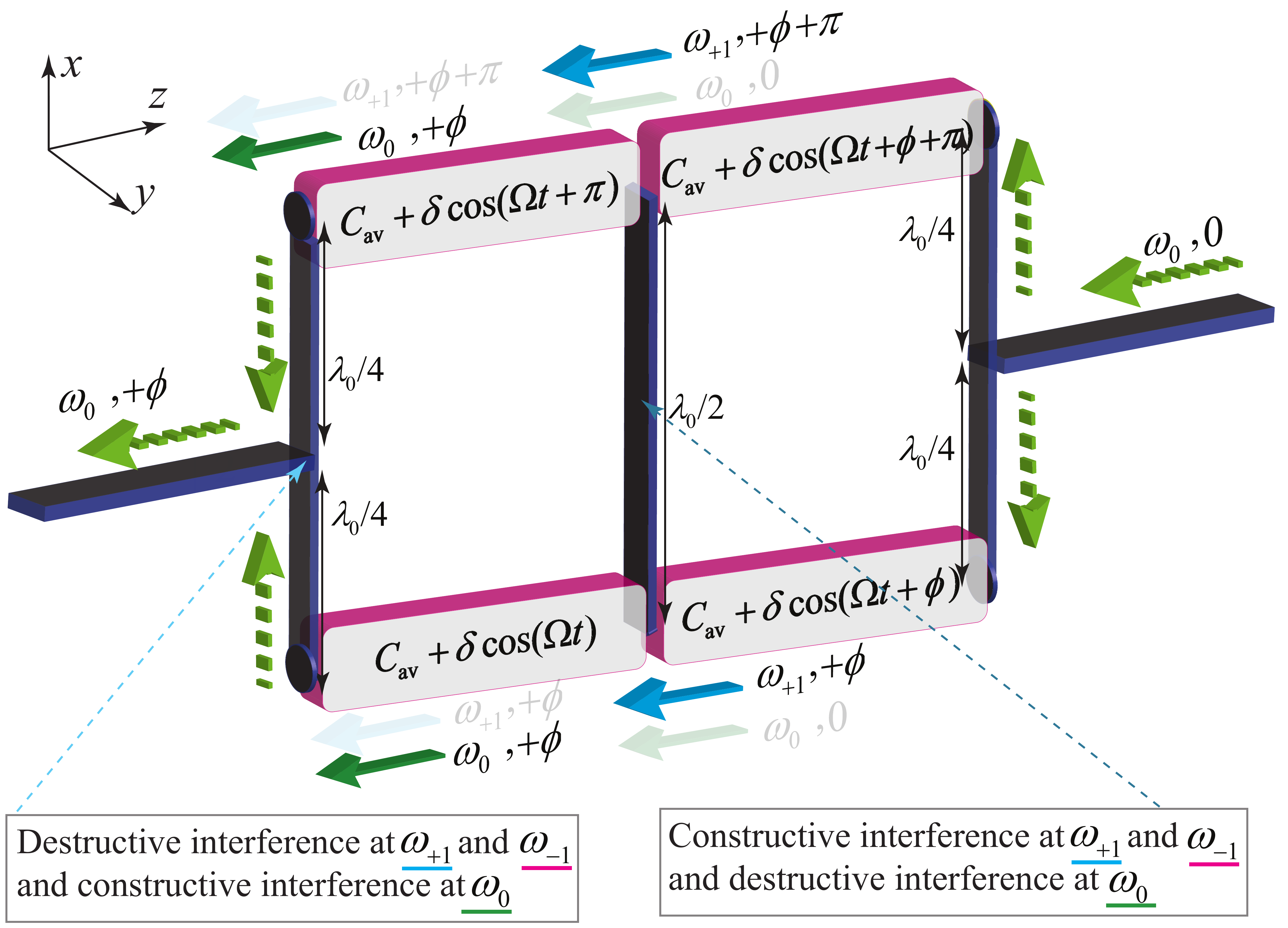}}
		\caption{Engineered frequency and phase transitions in the proposed temporal nonreciprocal phase shifter. (a) Forward signal injection. (b) Backward signal injection.} 
		\label{Fig:pp}
	\end{center}
\end{figure}

Figures~\ref{fig:p1} and~\ref{fig:p2} illustrate engineered frequency and phase transitions in the proposed temporal nonreciprocal phase shifter leading to opposite phase shifts for forward and backward signal transmissions. The structure is designed is a way to provide desired constructive and instructive interferences of different harmonics, and for both forward and backward signal transmissions. Two power splitters/combiners possessing two quarter-wavelength arms are utilized at two sides of the structure to attain the desired constructive and destructive interferences. In addition, a half-wavelength transmission line is placed at the middle of the structure to separate the two loops from each other and provide desired constructive and destructive interferences at different time harmonics. We aim to achieve a two-way complete signal transmission, but with opposite phase shifts for forward and backward transmissions, by proper engineering of the forward (Figures~\ref{fig:p1}) and backward (~\ref{fig:p2}) frequency and phase transitions in the proposed temporal nonreciprocal phase shifter. 

Consider the forward signal injection in Fig.~\ref{fig:p1}. The input signal is injected to the common port of the left power divider and flows in the upper and lower arms of the left loop. Next, these two signals make a transition to higher order time harmonic $\omega_{+1}$ but with the $+\pi$ and $0$ phase shifts in the upper and lower arms, respectively. Since the half-wavelength transmission middle interconnector makes a constructive interference at $\omega_{+1}$, the generated $\omega_{+1}$ time harmonic reaches to the right side loop. This leads to the generation of $\omega_0$ harmonic by the right side loop (second loop for the forward transmission), possessing identical phase shift of $-\phi$ in the upper and lower arms of the output of the right side loop. As a consequence, since the power divider is formed by equal quarter-wavelength long arms, the signals sum up at the common port of the right side power divider. This yields a transmission at $\omega_0$ with the phase shift of $-\phi$. We shall stress that, since the higher order time harmonics $\omega_{+1}$ and $\omega_{-1}$ acquire a $\pi$ phase shift difference at the two arms of the output (right) power divider, they will form a complete null at the output of the structure.

Now, consider the backward signal injection shown in Fig.~\ref{fig:p2}. The input signal is injected to the common port of the right side power divider and flows in the upper and lower arms of the right loop. Then, these two signals make a transition to higher order time harmonic $\omega_{+1}$ but with the $+\phi+\pi$ and $+\phi$ phase shifts in the upper and lower arms, respectively. Similar to the forward signal transmission case, the half-wavelength transmission middle interconnector makes a constructive interference at $\omega_{+1}$ and a destructive interference at $\omega_{0}$. Hence, the generated $\omega_{+1}$ time harmonic reaches to the left side loop. This leads to the generation of $\omega_0$ harmonic by the left side loop (second loop for the backward transmission), possessing identical phase shift of $+\phi$ in the upper and lower arms of the output of the right side loop. As a result, since the power divider is formed by equal quarter-wavelength long arms, the signals sum up at the common port of the left side power combiner. This results in a transmission at $\omega_0$ with the phase shift of $+\phi$ which is the opposite of the phase shift of the forward direction. Since the higher order time harmonics $\omega_{+1}$ and $\omega_{-1}$ acquire a $\pi$ phase shift difference at the two arms of the output (left) power divider, they will form a complete null at the output of the left power combiner. Such a contrast between the phase shift of the forward and backward transmissions originates from the asymmetric frequency-phase transitions in phase-engineered time-modulated structures.

\section{Experimental Design}

\begin{figure}
	\centering
	\includegraphics[width=0.9\columnwidth]{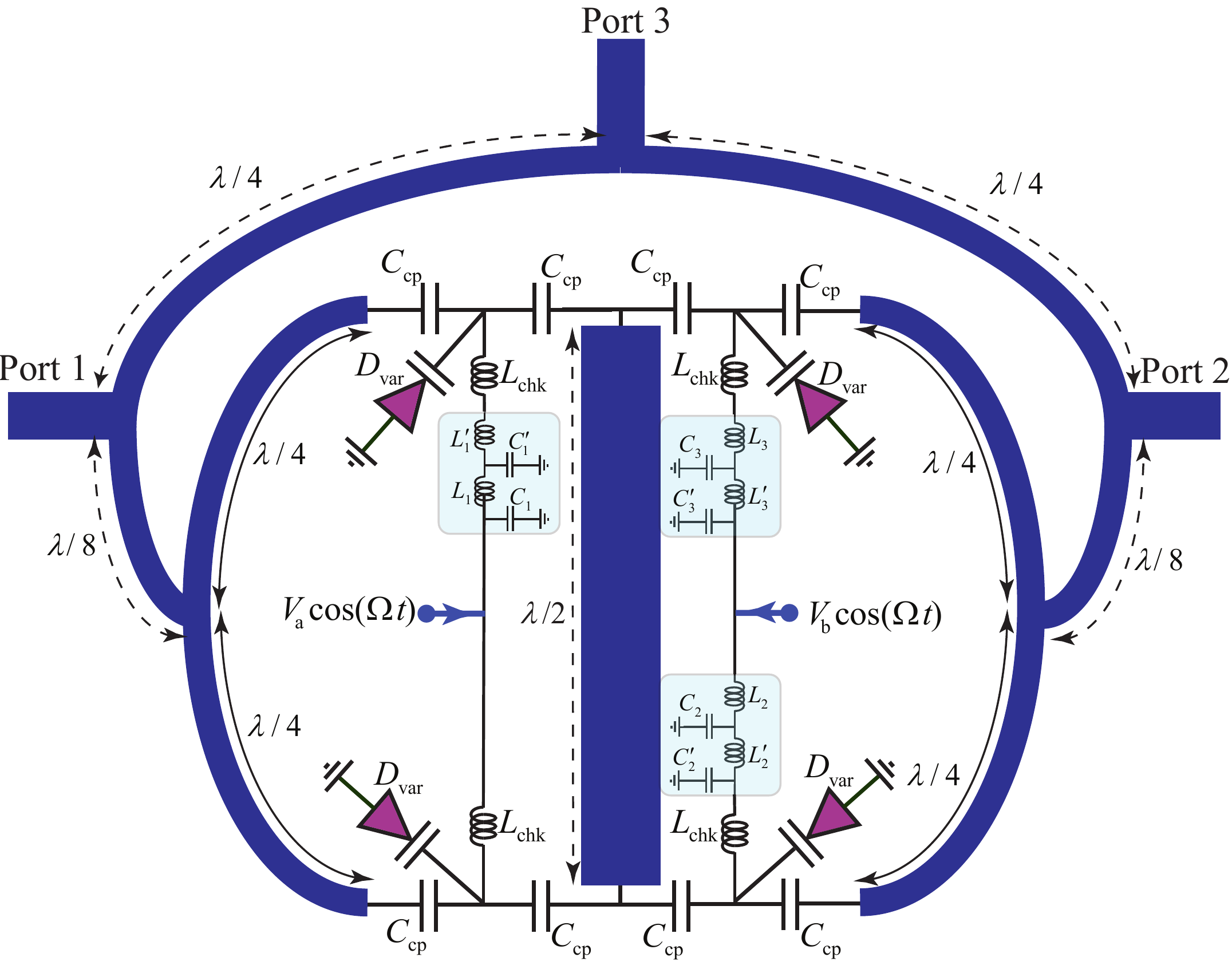}
	\caption{Implementation of the time-modulated circulator by four phase-engineered time-varying varactors.}
	\label{fig:2}
\end{figure}

Figure~\ref{fig:2} shows an implementation of the time-modulated circulator by four phase-shifted time-varying varactors. Here, four varactors are used for the creation of a phase-shifted time modulation. Two power splitters feed the four temporal transmission lines, and eight fixed capacitances are considered for decoupling of the DC bias and low frequency modulation signal from the incident and transmitted high frequency incident microwave signal.

\begin{figure}
	\centering
	\subfigure[]{\label{fig:ph_front} 
		\includegraphics[width=60mm]{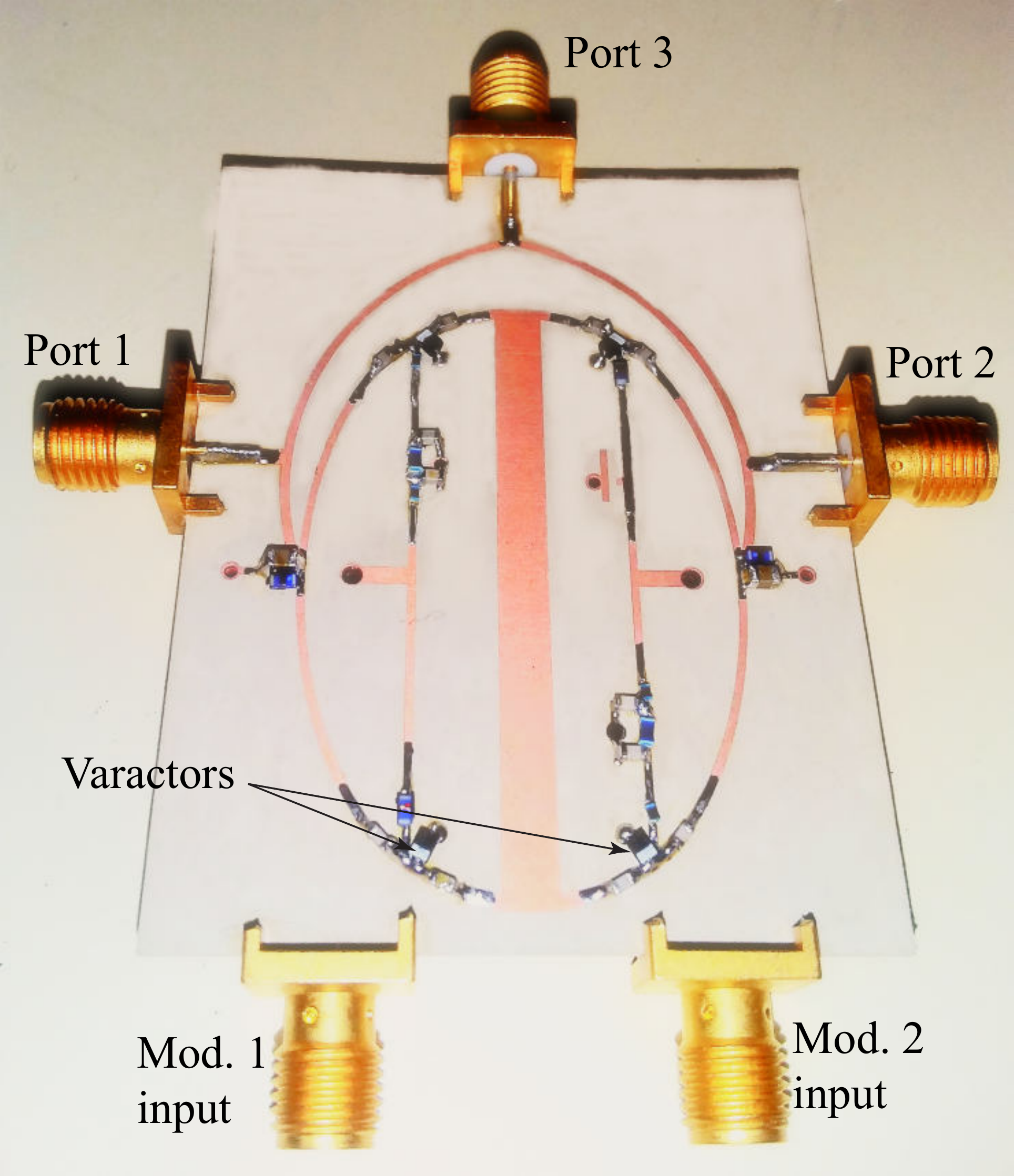}}
	\subfigure[]{\label{fig:ph_back}  
		\includegraphics[width=60mm]{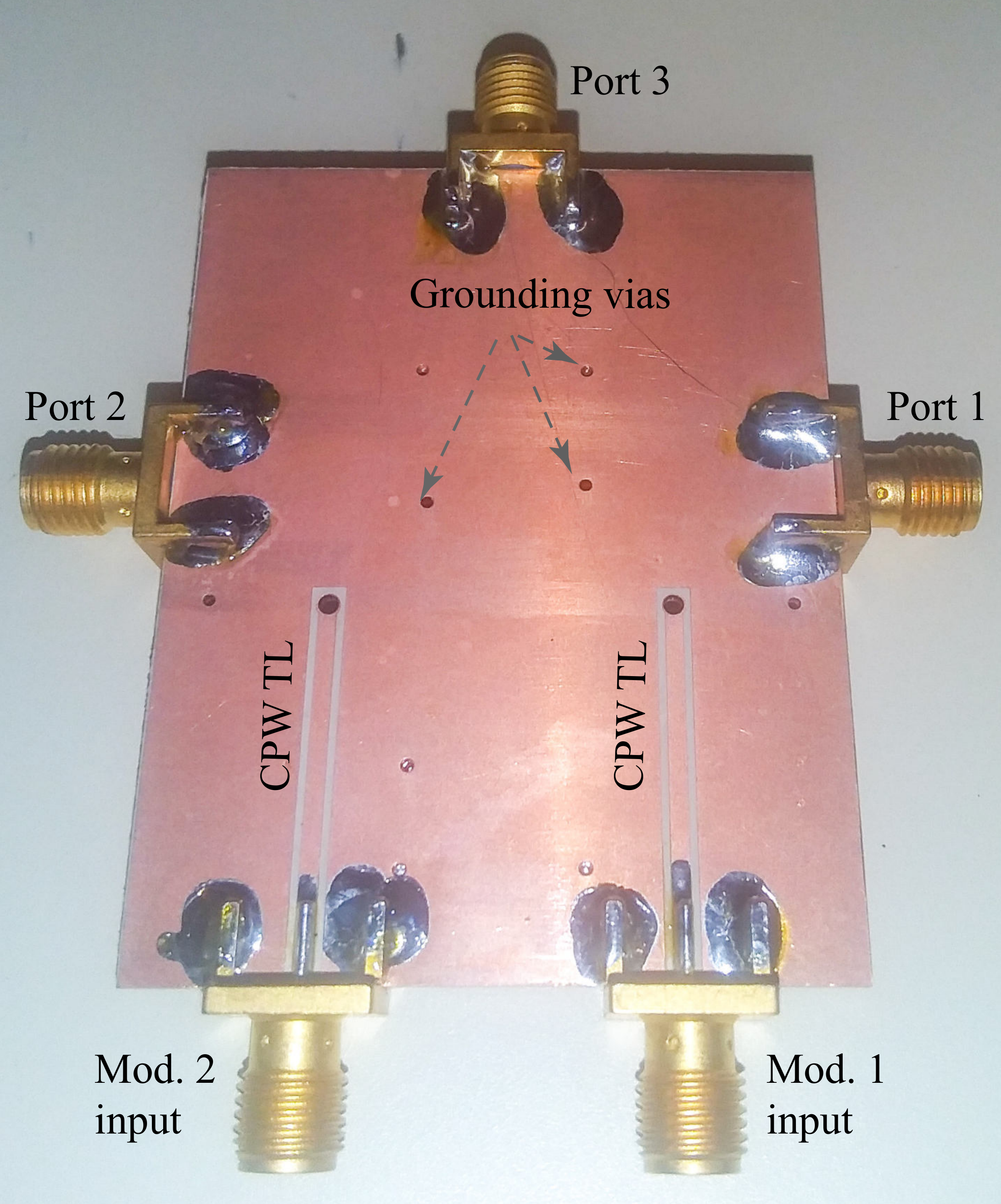}}	
	\caption{A photo of the fabricated time-modulated circulator. (a) Top view. (b) Bottom view.}
	\label{fig:3}
\end{figure}

Figures~\ref{fig:ph_front} and~\ref{fig:ph_back} show the top and bottom views of the fabricated circulator, respectively. The circulator is designed at the center frequency of 1.56 GHz with the modulation frequency of $\Omega=101.1$ MHz and modulation power of 15 dBm. Four BB837 varactors manufactured by Infineon Technologies have been used for realizing the time modulation. We have utilized a RT6010 substrate with permittivity $\epsilon_\text{r} = 10.2$, thickness $h = 50$~mil and $\tan \delta = 0.0023$. Port 1-3 are the main high frequency ports of the circulator (support 1.56 GHz), while two other ports at the bottom of the structure, i.e., Mod. 1 input and Mod. 2 input, support the low frequency modulation signal injection with different phase shifts. The two 50 Ohm coplanar waveguide (CPW) transmission lines at the back of the structure, shown in Fig.~\ref{fig:ph_back}, possess a width of W=45 mils and a gap of G=20 mils. We have used four BB837 varactors manufactured by Infineon Technologies for realizing the phase-shifted time modulation. We utilized two ZFBT-6GQ+ minicircuits bias-tees.

\begin{figure}
	\begin{center}
		\subfigure[]{\label{fig:meas_su} 
			\includegraphics[width=80mm]{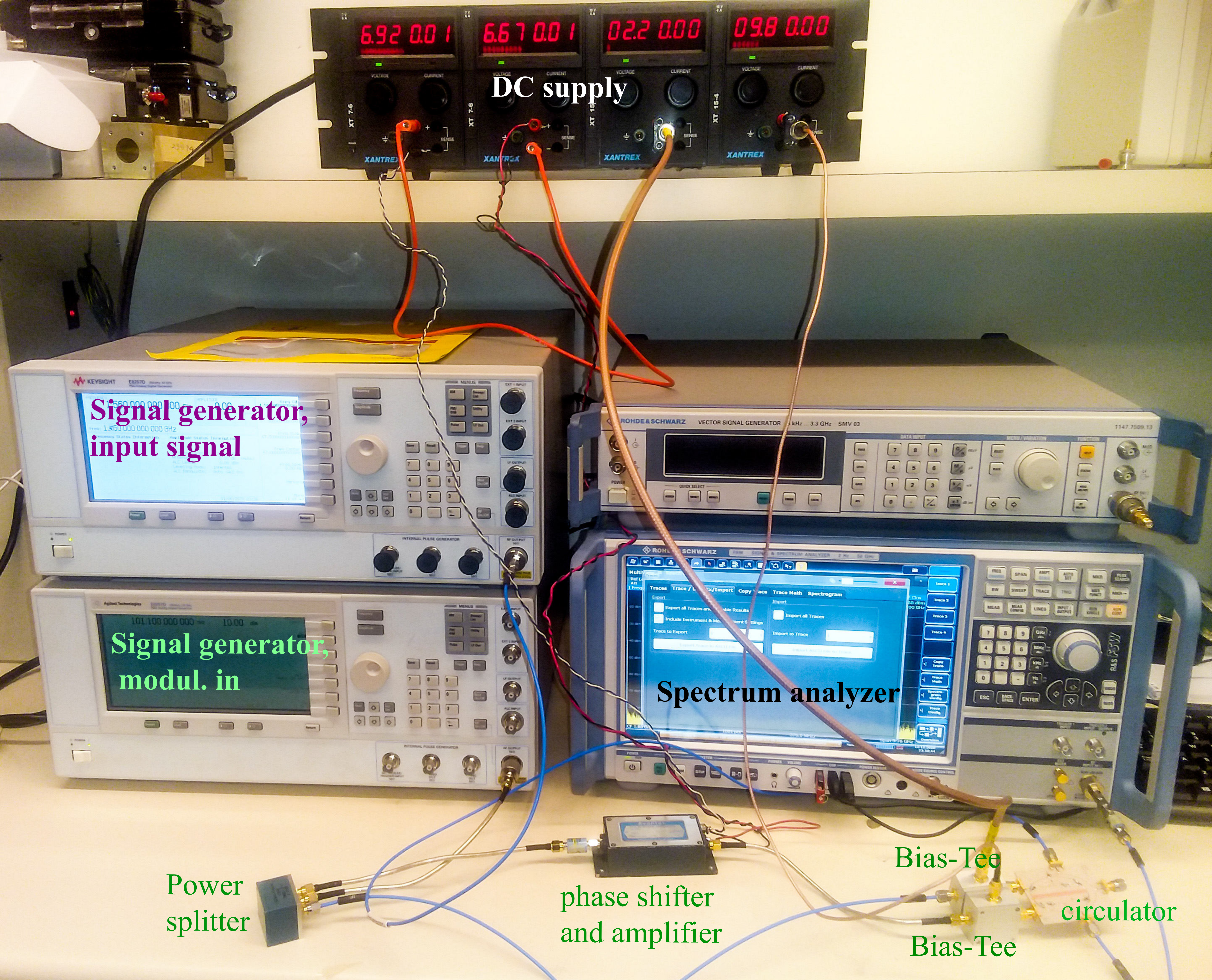}}					
		\subfigure[]{\label{fig:measf} 
			\includegraphics[width=0.55\columnwidth]{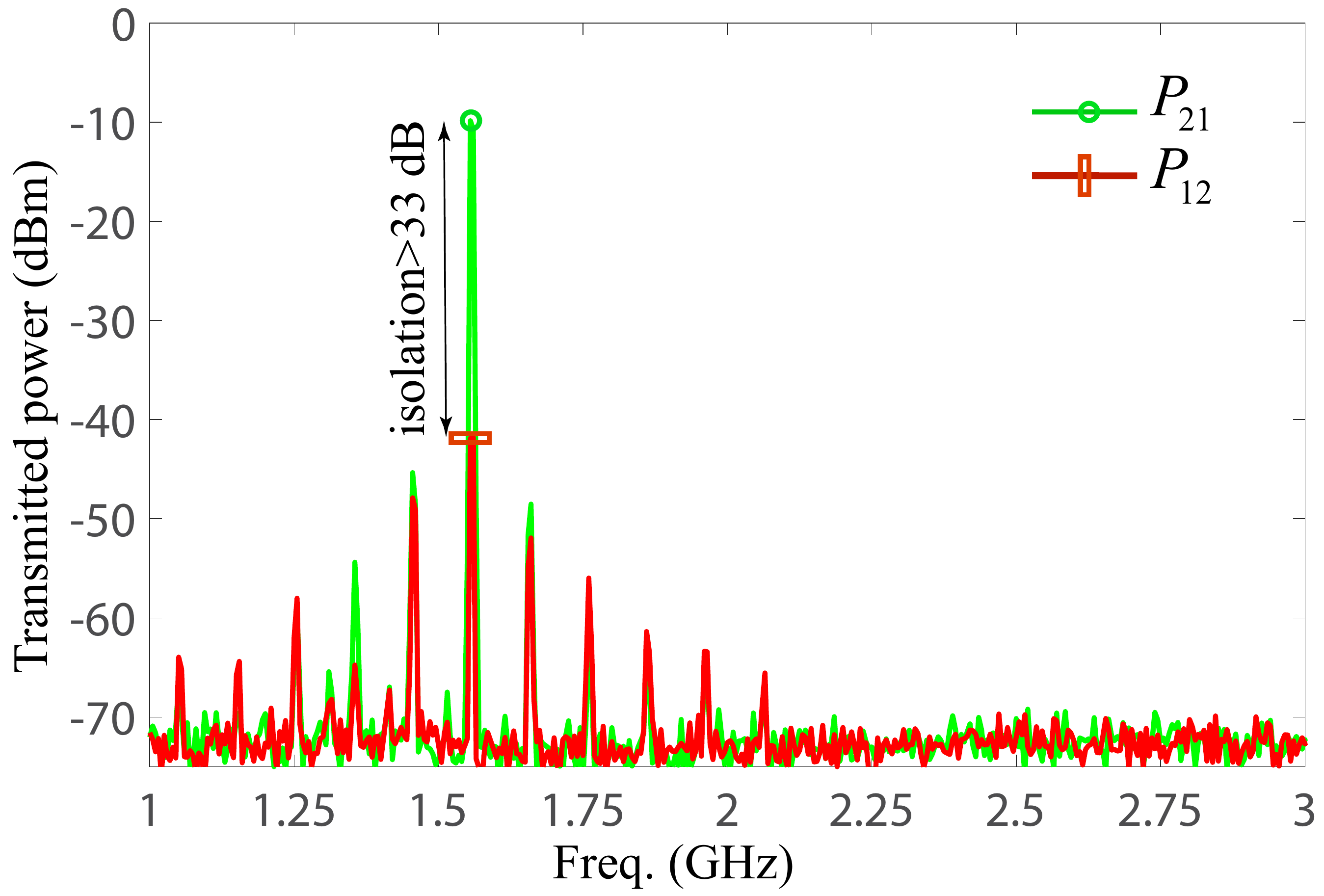}} 
		\subfigure[]{\label{fig:measb} 
			\includegraphics[width=0.55\columnwidth]{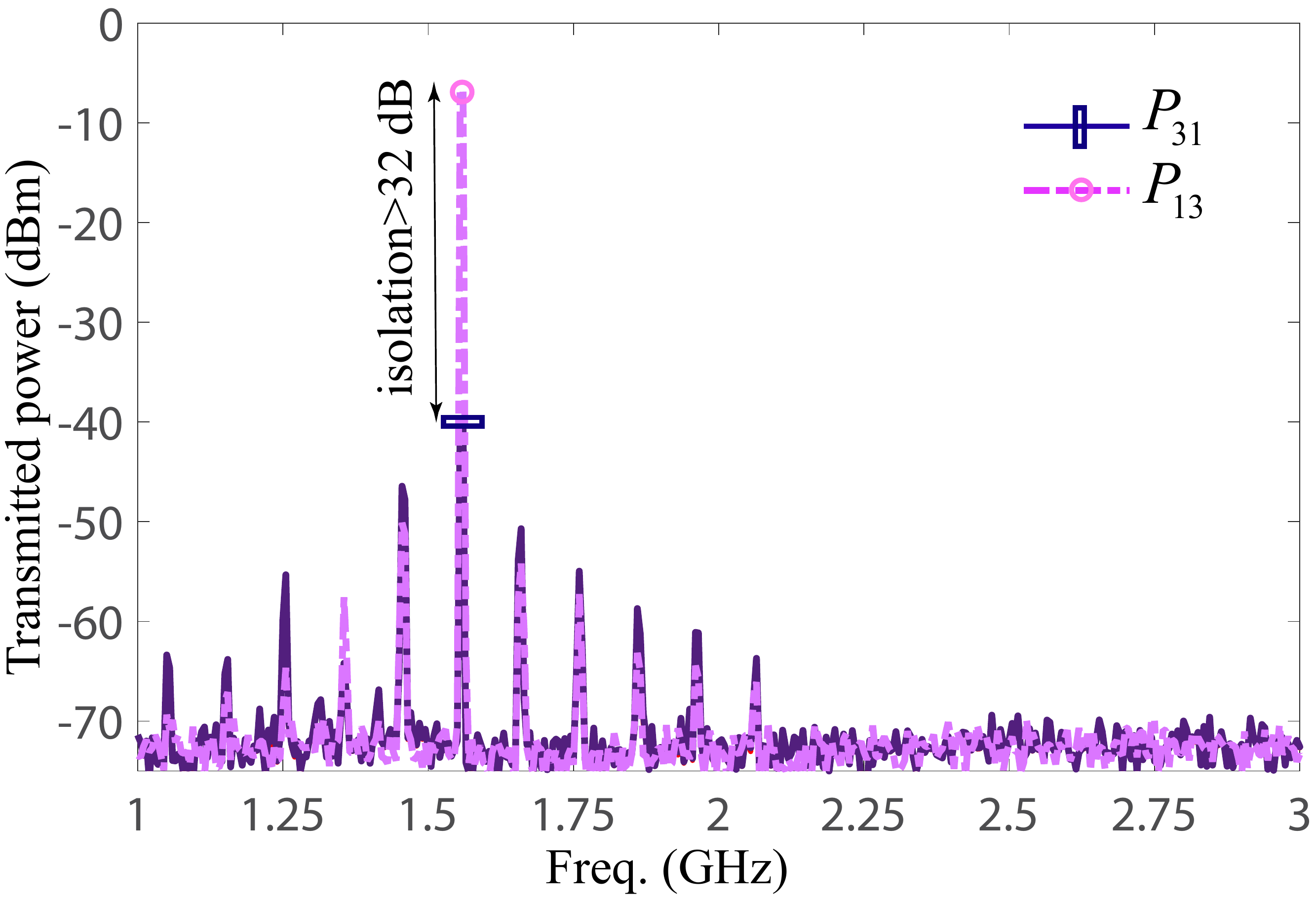}}
		\caption{Experimental results. (a) A photo of the measurement set up. (b) Power transmission from port 1 to port 2, $P_{21}$, and from port 2 to port 1, $P_{12}$. (c) Power transmission from port 1 to port 3, $P_{31}$, and from port 3 to port 1, $P_{13}$.} 
		\label{Fig:meas}
	\end{center}
\end{figure}

Figure~\ref{fig:meas_su} shows a photo of the experimental set-up for the measurement of the unidirectional signal transmission through the temporal circulator, where $\Omega=2\pi\times101.1$~MHz. Figures~\ref{fig:measf} and~\ref{fig:measb} plot the experimental results for wave circulation between different ports of the circulator. It may be seen from the experimental results that an isolation of more than 32 dB between the ports is achieved, whereas the undesired time harmonics are highly suppressed, they are more than 40 dB lower than the main signals.

\section{Discussions}

Table~\ref{Tab:1} summarizes the performance of the fabricated circulator and provides a comparison between the performance of this circulator and that of recently proposed circulators with different techniques, including space-time (ST) modulation, magnetic-ferrite-loaded waveguides, nonlinearity, and transistor-loaded transmission lines. In this table, IL stands for insertion loss, RL stands for return loss, and FBW is an acronym for the fractional bandwidth which is equal to the frequency bandwidth over the center frequency. Additionally, P1dB is the output 1 dB compression point which shows the prominent linearity of the circulator. The proposed circulator can be realized in a much more compact fashion, and may be integrated into chip technology for high frequencies thanks to the availability of variable capacitors at high frequencies~\cite{lira2012electrically}. In addition, in contrast to the transistor-based circulator~\cite{Tanaka_1965} where nonlinear and low power handling transistors are placed in series with the incident signal, our circulator is endowed by high power rating as the varactors are placed in parallel to the incident signal. In our experiment, we applied up to +33 dBm input signal, and the power rating is expected to exceed $47$~dBm. Having leveraged the linear properties of the time modulation, the proposed temporal circulator exhibits an outstanding linear response, where the P1dB is +31.7 dBm and IIP3 is +45.4 dBm.

	\begin{table}
		\small
		\caption{Summary of the efficiency of the time-modulated circulator and comparison with previously reported circulators.}
		\medskip
		\centering
		\begin{tabular}{|>{\centering\arraybackslash}m{0.8cm}|>{\centering\arraybackslash}m{2.8cm}| >{\centering\arraybackslash}m{1.7cm}|>{\centering\arraybackslash}m{1.0cm}|>{\centering\arraybackslash}m{1.6cm}|>{\centering\arraybackslash}m{0.85cm}|>{\centering\arraybackslash}m{1.5cm}|>{\centering\arraybackslash}m{0.8cm}|>{\centering\arraybackslash}m{0.85cm}|>{\centering\arraybackslash}m{0.9cm}|}
			\hline
			  & Technology&  Area & \shortstack{Freq.\\(GHz)}& FBW (Isolation)& \shortstack{Isol.\\ (dB)}& \shortstack{IL\\ (dB),\\ IL$_\text{21}$/IL$_\text{13}$ }&\shortstack{ RL\\ (dB)} &\shortstack{ P$\text{1dB}$\\(dBm)} &\shortstack{ IIP3\\(dBm)} \\
			\hline
			This work &  \shortstack{Phase-engineered \\time modulation }& \shortstack{$0.067\lambda^2$\\ $(2475 \text{mm}^2)$}& 1.56 & 19.2$\%$ (20~dB)& 32 & 7.2/9.1 & 16.2 & 31.7& 45.4\\ \hline
			\cite{martinez2019low}  & Ferrite magnets & \shortstack{$0.8\lambda^2$\\ $(500 \text{mm}^2)$} & 12 &  33.3$\%$ (10~dB)& 10 & 1 & 10 & N/A& N/A\\\hline
			%
			%
			\cite{kord2017magnet} & Time modulated resonators & \shortstack{$0.006\lambda^2$\\ $(575 \text{mm}^2)$}  & 1 &  4$\%$ (20~dB) & 55 & 3.3  &10.8 &29& 33.7\\ \hline
			\cite{qin2016broadband}  & Chip-space-time modulation & \shortstack{$0.0003\lambda^2$\\ $(11.76 \text{mm}^2)$}   & 1.6 &  18$\%$ (25~dB) & 25 & 2 & $>10$& 10.5& N/A \\\hline
			\cite{dinc2017millimeter}  & Chip-space-time modulation across delay line &\shortstack{$0.015\lambda^2$\\ $(2.16 \text{mm}^2)$} & 25 &  18$\%$ (18.5~dB) & 18.5 & 3.3/3.2 &N/A & 21.5 & 19.9\\\hline			
			\cite{jain20180}  & Chip-hybrid-coupler N-path & \shortstack{$0.000006 \lambda^2$\\ $(1\text{mm}^2)$}& 0.725 & 27$\%$ (30~dB) & 30 & 3.1/15 & N/A & +5.5& 14 \\\hline				
			\cite{nagulu2020multi}  & Chip-switched TL gyrator &\shortstack{$0.0001 \lambda^2$\\ $(9\text{mm}^2)$}& 1 &18.4$\%$ (25~dB)& 25 & 2.07/2.49 & N/A &17.3 & 36.5\\\hline			
			\cite{yang201885}  & \shortstack{Chip-transistor\\ (65-nm CMOS),\\Delta resonators} & \shortstack{$0.023 \lambda^2$\\ $(0.21 \text{mm}^2)$}  & 100 &1.5$\%$ (45~dB) & 46 & 5.6 & $>10$& 11.4 & N/A\\\hline			
			\cite{wu2010cmos}  & \shortstack{Chip-transistor\\ (180-nm CMOS),\\Feedforward} & \shortstack{$0.021 \lambda^2$\\ $(3.22 \text{mm}^2)$} & 24& 1$\%$ (15~dB)& 20 & -12.3/ -22.4 (gain)& $>8$ & -19.8& -11\\\hline
			\cite{chang201030}  & \shortstack{Chip-transistor\\ (180-nm CMOS),\\Current reuse} & \shortstack{$0.0036 \lambda^2$\\ $(0.36 \text{mm}^2)$} & 30 &7.6$\%$ (15~dB) & 12 & 7.5/5 & $>6$ &  -7& N/A\\\hline
			\cite{huang201224}  & \shortstack{Chip-transistor\\ (180-nm CMOS),\\Cancellation} & \shortstack{$0.0022 \lambda^2$\\ $(0.35\text{mm}^2)$} & 24 &11.2$\%$ (15~dB) &30  & 8.5/9 & $>8.5$ &-12 & N/A\\\hline
			\cite{hung2013ultra}  & \shortstack{Chip-transistor\\ (90-nm CMOS),\\Amps} & \shortstack{$0.01 \lambda^2$\\ $(0.93\text{mm}^2)$} &31.75 &58$\%$ (15~dB) & 12$\sim$22 & 9/3.5 &$>3$ & -3 & N/A\\\hline
			\cite{Wang_TMTT_2015}  & \shortstack{Chip-transistor\\ (180-nm CMOS),\\isolators} & \shortstack{$0.004 \lambda^2$\\ $(0.715 \text{mm}^2)$} & 24 & 3$\%$ (15~dB) & 20 & 5.7/5.7 & $>16$ &9.5  & N/A\\\hline
		\end{tabular}
		\label{Tab:1}
	\end{table}

\section{Conclusion}
In summary, we have proposed a lightweight compact non-magnetic time-varying circulator. The circulator is based on a time-modulated nonreciprocal phase shifter. The proposed nonreciprocal phase shifter exhibits great performance including nonreciprocal phase shift range, and can be adopted for broad-band operation. The experimental results demonstrate a strong isolation level of more than 32 dB which can be further improved. Furthermore, the unwanted time harmonics are 40 dB lower than the main signal. More importantly, this circulator exhibits a highly linear response, with the P1dB of +31.7 dBm and the IIP3 of 45.4 dBm. Such promising results and compact profile of the circulator can be further improved and optimized in future work. Furthermore, the proposed structure can be redesigned for integration into integrated circuit technology to achieve much smaller profile and versatile operation.

{
\textbf{Author contributions:}\\
S.T. carried out the analytical modeling, numerical simulations, sample fabrication, and measurements. G.V.E. planned, coordinated, and supervised the work. All authors discussed the theoretical and experimental aspects and interpreted the results. All authors contributed to the preparation and writing of the manuscript. Correspondence and requests for materials should be addressed to Sajjad Taravati~(email: sajjad.taravati@utoronto.ca).
}

\textbf{Notes}\\
The authors declare no competing interests.

\begin{acknowledgement}

This work is supported by the Natural Sciences and Engineering Research Council of Canada (NSERC). 

\end{acknowledgement}


\bibliography{Taravati_Reference}

\end{document}